# Observation of large perpendicular magnetic anisotropy and excessive polar magneto-optical effect in Pt/CoFeB/Ru tri-layer system

Md Rejaul Karim[1], Aman Agrahari[1], Arun Singh Dev[1], Rakhul Raj[2], Mohd S Sabir[1], Arun Jacob Mathew[4], Joseph Vimal Vas[3], Yasuhiro Fukuma[4,5], V. Raghavendra Reddy[2], Rohit Medwal[1*]

[1]Department of Physics, Indian Institute of Technology Kanpur, Kanpur 208016, India

[2]UGC-DAE Consortium for Scientific Research, University Campus, Khandwa Road, Indore 452001, India

[3]The Ernst Ruska-Centre for Microscopy and Spectroscopy, Forschungszentrum Jülich 52428, Germany

[4]Department of Physics and Information Technology, Faculty of Computer Science and Systems Engineering, Kyushu Institute of Technology, Iizuka 820-8502, Japan

[5]Research Center for Neuromorphic AI hardware, Kyushu Institute of Technology, Kitakyushu 808-0196, Japan

[*]Correspondence should be addressed to Rohit Medwal: rmedwal@iitk.ac.in

**Abstract:**

Heterostructures comprising ferromagnet (FM) and heavy metals (HM) with perpendicular magnetic anisotropy (PMA) and interfacial Dzyaloshinskii-Moriya interaction (iDMI) can host chiral domain walls and topological spin textures, making them highly promising for various spintronics applications. In this paper, we have investigated the magneto-optical properties, the anomalous Hall effect (AHE), and PMA of Pt/CoFeB/Ru multilayers engineered to possess significant iDMI. We utilized the Anomalous Hall effect (AHE), and the polar magneto-optical Kerr effect (p-MOKE), Hall response and the domain wall motion in Pt/CoFeB/Ru-systems. Both MOKE and AHE measurements confirm that the films maintain strong perpendicular magnetization for CoFeB thicknesses below 1.2 nm. The effective magnetic anisotropy $K_{eff}$ of $0.88 \times 10^6$ erg/cm$^3$ has been achieved without any post-annealing, highlighting the high-quality interface in this multilayer design. The angular dependence of the switching field deviates from the conventional Kondorsky model and is well described using a modified Kondorsky formalism, capturing the role of field-induced domain-wall softening and pinning effects in the reversal process. Furthermore, p-MOKE microscopy imaging during the magnetization reversal process provides detailed insight into domain nucleation and subsequent domain-wall propagation. The observations reveal well-defined, stable magnetic domains that evolve coherently under the applied magnetic field. Such a behavior is expected in the system where interfacial DMI, PMA interact to stabilize the chiral Néel-type domain walls, which are essential for fast, low-power domain-wall motion driven by spin-orbit torques.

**Introduction:**

The spin-orbit coupling (SOC)-related antisymmetric exchange interaction and the interfacial Dzyaloshinskii-Moriya interaction (iDMI) leads to the formation of domain walls with a chiral spin arrangement[1–4]. When the strength of the iDMI surpasses a critical threshold, it can stabilize skyrmions, which are topologically protected spin structures in magnetic multilayers[5–8]. These topologically protected spin textures are expected to be used in future magnetic memories, sensors, all optical helicity dependent switching and computing devices, etc. Since the discovery of magnetic skyrmions in 2009[9], significant theoretical and experimental efforts have been put in exploring the skyrmions driven by iDMI and PMA in multilayer thin films. The interfacial iDMI arises from the strong spin-orbit coupling of a heavy metal (HM) in contact with an ultrathin ferromagnetic material (FM), where the surface's broken inversion symmetry plays a crucial role[10]. The required symmetry reduction can be established at the interface of a bilayer comprising FM and HM films, where symmetry reduction is achieved at the interface and described by Fert's triangle[11], consisting of two magnetic atoms and a spin-orbit coupling (SOC)-carrying atom on the heavy metal side. The asymmetry of bubble domain expansion in an in-plane (IP) field has been used as a common indicator in experimental studies of DMI across various CoFeB and Co-based samples, which is used to estimate the strength of the DMI energies in the range of 0.5-2.5 mJ/m$^2$ for systems like Ta/CoFeB/MgO[12], CoFeB/MgO[13], Pt/Co/Ir/Pt[14], Pt/Co/Ir[15]. On the other hand, PMA is observed in multilayers due to the broken inversion symmetry, which plays a vital role for enhancing thermal stability[16], enhancing device endurance and reducing critical current densities in spin-transfer torque magnetic random-access memories (STT-MRAM)[17]. Numerous thin films with PMA have been studied for spintronics application such as, CoFeB/MgO[18,19] films, CoFe/Pd[20], Co/Pt[21], and single-layer films like CoPt[22], MnGa[23], and FePt[24], etc. Among them, CoFeB-based films[25] are particularly appealing due to their high spin polarization and moderate saturation magnetization, which are crucial for lowering the switching current density.

In parallel with the ongoing research aimed at increasing iDMI and PMA, it is also important to engineer multilayer stacks to optimize chiral interaction, domain wall dynamics, magnetic anisotropy, etc. As the magnitude and sign of DMI is dependent on the SOC of heavy metal and ferromagnetic thickness, we could control the domain chirality and enhanced the stability of magnetic textures. The Ru layer also offers the way to modulate the interfacial hybridization, influence magneto-elastic and Ruderman–Kittel–Kasuya–Yosida (RKKY) coupling across the stack. So, the multilayers like Pt/CoFeB/Ru are engineered to exhibits chiral spin structures while maintaining strong magneto-optical response and desirable transport properties. In our study, we thoroughly investigated Pt/CoFeB/Ru structures, combining with anomalous Hall effect (AHE) measurements, polar MOKE characterization, and direct Kerr microscopy imaging. The effective anisotropy and domain behavior observed in our samples highlights the pivotal role of interface engineering in tuning both the magnitude of PMA and the interfacial contributions linked to DMI.

I. **Experimental methods:**

The multilayer stacks were deposited on a Si/SiO$_2$ substrate by the magnetron sputtering method at room temperature. The stacks consisted of [Pt (2.5 nm)/Co$_{60}$Fe$_{20}$B$_{20}$ (t nm)/Ru (0.6 nm)] where t= 0.6, 0.9, 1.2, 1.4, 2 nm and [Pt (2.5 nm)/Co$_{60}$Fe$_{20}$B$_{20}$ (0.9 nm)/Ru (0.6 nm)]$_x$, where x= 1, 2, 5, 7. Prior to the Pt deposition, Ta (4 nm) seed layer was deposited on a thermally oxidized Si substrate, followed by the trilayer of Pt/CoFeB/Ru. A Pt capping layer (2 nm) was deposited to protect the Ru layer from oxidation and degradation due to exposure to the atmosphere. The sputtering system maintained a base pressure below 5×10$^{-8}$ mbar before the deposition. In this system, the Pt (bottom layer) induces both PMA and DMI, while the Ru spacer layer is believed to induce only perpendicular magnetic anisotropy.

A vibrating sample magnetometer (VSM) was used to measure the static magnetic properties of the samples with the magnetic field applied perpendicular to the sample plane, allowing for the determination of the saturation magnetization ($M_s$). The thickness and interface roughness of the samples were measured using X-ray reflectivity (XRR) with a Co-K$_\alpha$ x-ray source of wavelength 1.789 Å. The p-MOKE[26] hysteresis loops were recorded using a MOKE microscope (M/s Evico Magnetics, Germany) featuring a broadband white LED light source. All measurements were conducted with a 5x objective lens. Importantly, the MOKE microscope employs a differential subtraction technique to acquire domain images, subtracting a previously captured optical image from the live optical image. A set-up with a 1.0-T electromagnet, a Keithley 6221, and a Keithley 2182A was used to study the anomalous Hall effect (AHE) of all fabricated devices.

**Results and Discussion:**

The out-of-plane (OOP) magnetic hysteresis loops were measured by using the magneto-optical Kerr effect (MOKE) with an external field up to ±3000 Oe at room temperature. Fig. 1(b)-(d) shows the full loop for CoFeB with thicknesses of 0.6, 0.9, and 2nm respectively, where it is clear that 0.9nm thickness sample has a large hysteresis loop and 2 nm sample does not show OOP magnetization. To explain this, we can conclude that PMA generally originates from broken symmetry and hybridization effects at the interface depending on the system. For the 0.6nm film, the thickness is very low, close to the limit of continuity for CoFeB[27], so the PMA is present but not fully developed due to possible island-like growth and a partial effect of magnetic dead layer. At 0.9nm film, the layer is expected to be continuous and the interfacial PMA dominates over the demagnetizing energy. The effective magnetic anisotropy energy density ($K_{eff}$) and its relationship with FM thickness is important to the magnetic behavior of the material. $K_{eff}$ is defined by the $K_{eff} = H_k M_s / 2$, where, $M_s$ is saturation magnetization measured by VSM measurement and $H_k$ is the defined field required to switch the magnetization direction. The interfacial anisotropy ($K_i$) was evaluated by plotting the thickness dependence of anisotropic energy densities for the deposited samples ($K_{eff}.t_{CoFeB}$) as a function of the $t_{CoFeB}$ layer as shown in Fig. 1(e). At greater thicknesses, the anisotropy energy is negative, indicating a preference for in-plane anisotropy, with magnetic moments aligned with the film's plane. This behavior is expected, as the demagnetizing energy $2\pi Ms^2$ dominates in a thicker ferromagnetic layer and suppresses perpendicular anisotropy. As the thickness of CoFeB decreases, the anisotropy energy becomes less negative, reaching a maximum positive value before reducing again at a lower thickness. The maximal value of ($K_{eff}.t_{CoFeB}$) (0.081 erg/cm²) is observed at an ideal FM layer thickness of 0.9 nm, where the heterostructure has the highest PMA energy density. The equation to calculate the $K_i$ is given by, $K_{eff}\, t_{CoFeB} = (K_v - 2\pi Ms^2)t_{CoFeB} + K_i$ and the value of $K_i$ was obtained as 0.23 erg/cm². This value is small compared to the conventional CoFeB/MgO interface[28,29], where strong p-d hybridization enhances interfacial PMA, but larger compared to the other reported CoFeB/Pt interface[30,31] where intermixing and spin-orbit scattering generally reduce interfacial anisotropy. It is worth mentioning that the volumetric contribution is negative, confirming that the bulk favors an in-plane easy axis. The absolute value of the $K_v$ is larger compared to the bulk-Fe ($K_v=0.45\times10^6$)[32] but less than the hcp-Co ($K_v=5.3\times10^6$)[33]. This intermediate value reflects the mixed Co-Fe composition and the structural ordering of the CoFeB alloy. Overall, the analysis confirms that the PMA in the system is primarily interfacial in origin, with the strongest perpendicular anisotropy occurring at a CoFeB thickness of 0.9 nm.

Atomically resolved high angle annular dark-field (HAADF) scanning transmission electron microscopy (STEM) was used to investigate the interface quality, composition, and elemental distribution of the layers. The samples were prepared using a Zeiss Crossbeam focused ion beam (FIB) with a Pt layer on the top to protect the sample from electron beam. Fig. 1(f) presents the energy-dispersive X-ray (EDX) elemental maps acquired simultaneously with STEM imaging. The maps clearly reveal the spatial distribution of each constituent element across the heterostructure, confirming sharp, well-defined interfaces and the exact intended layer sequence[34]. The lack of noticeable interdiffusion or elemental intermixing between adjacent layers demonstrates the excellent structural integrity of the stack and validates the optimized growth conditions employed during fabrication.

The magnetic properties of the samples were measured through AHE, which is a widely used technique for characterizing perpendicularly ferromagnetic thin films. The samples were patterned into a Hall bar with dimensions of 10 μm × 50 μm (Fig. 2a) using an optical lithography and lift-off process. Small pads at each end of the Hall cross were subsequently metallized with thick Ti/Au electrodes to provide reliable electrical contacts. The Hall measurements were carried out at RT with an external magnetic field applied perpendicular to the sample plane and the excitation current and measuring voltage are perpendicular to each other. Fig. 2(b) shows the hysteresis loops for the sample with Pt (2.5)/CoFeB (0.9 nm)/Ru (0.6)/Ta (1.5) for which, we observed a strong and robust formation of PMA from MOKE and VSM measurements. In the AHE measurement, the perpendicular anisotropy field ($H_k$) can be estimated by applying an external field ($H_x$) parallel to the plane of the film (Fig. 2(c)). Applying ($H_x$) gradually rotates the magnetization from the out-of-plane to the in-plane direction, leading to a reduction in the Hall resistance ($R_{xy}$). Based on the Stoner-Wohlfarth model ($H_k$) can be determined by fitting the relationship between ($R_{xy}$)/($R_{xy}^{max}$) Vs ($H_x$) within a small field range in the equation[28],

$$R_{xy}/R_{xy}^{max} = cos\,[arcsin(H_x/H_k)] \approx 1 - 0.5 \times (H_x/H_k)^2 \qquad (1)$$

where, $R_{xy}^{max}$ is the maximum Hall resistance when the external magnetic field is along the z-axis. As shown in the Fig. 2(c), the value of $H_k \sim 4.4$ kOe is obtained for the Pt/CFB (0.9 nm)/Ru multilayer. To investigate the magnetization switching mechanism, the switching field $H_{sw}$ was extracted from AHE measurements under an external magnetic field applied at different polar angles $\theta_H$. The angular dependence of $H_{sw}$, shown in Fig. 2(d), deviates from the conventional Kondorsky model[35,36] $H_{sw} \propto 1/\cos \theta_H$. Instead, $H_{sw}$ remains nearly constant and equal to $H_{sw}(0)$ up to $\theta_H \approx 45°$, followed by a rapid increase at larger angles. This domain wall-mediated reversal behavior can be understood by considering in-plane-field-induced domain-wall softening[37]. As $\theta_H$ increases, while the out-of-plane component that drives wall motion decreases, the in-plane field component progressively increases and reduces the domain-wall energy and effective pinning barriers by modifying the internal wall structure. Up to a critical angle, the reduction in the depinning barrier compensates for the loss of the out-of-plane driving field, resulting in a switching field insensitive to the field angle. Beyond this angle, domain-wall softening saturates as the wall magnetization reaches its equilibrium configuration, and the decreasing out-of-plane field component dominates, leading to a sharp rise in $H_{sw}$, similar to the Kondorsky model. To account for this deviation, the data are fitted using a modified Kondorsky relation,

$$\frac{H_{sw}(\theta_H)}{H_{SW}(0°)} = 1 + \left(\frac{\cos(t)}{\cos(\theta_H)} - 1\right) \times S \qquad (2)$$

Where $S = \frac{1}{1+e^{-(\theta_H - t)/\Delta}}$, is the sigmoid function, $\Delta$ is the transition width fixed at 1°, and $t$ is the fitting parameter which quantifies the threshold angle up to which the switching field remains insensitive to the field angle. Using this model, a value of $t = 42°$ is obtained for the single-stack Pt/CoFeB/Ru multilayers. This ability to maintain a nearly constant switching field at lower field angles is advantageous for achieving reliable and angle-tolerant magnetization switching in practical device applications.

To study how the number of multilayer repetitions affects PMA, samples with stack numbers x=1,2,5, and 7 were fabricated, and their AHE loops are shown in Fig. 3(a)-(d). The AHE results clearly reveal that the sample with x=2 stacks exhibit the strongest PMA, as evidenced by a square and well-defined hysteresis loop, making this stack particularly suitable for domain-imaging experiments such as MOKE microscopy. The enhanced PMA for the sample with 2 stacks could be due to an optimal balance between interfacial anisotropy and total film thickness make sure the interfacial anisotropy dominates over thickness-induced demagnetizing effects giving the strongest PMA among all the samples. For x=1, the limited number of interfaces can result in a weaker interfacial contribution to the effective anisotropy, on the other hand increasing the number of stacks to x = 5 or 7 introduces greater interfacial roughness, structural disorder, and atomic intermixing, which diminishes the effective PMA of the system.

Now to understand, the nucleation field ($H_n$), magnetization switching field ($H_{sw}$), we fabricated the Hall bars of dimensions 10 µm × 50 µm. The schematic of the device is shown in Fig. 4a. The anomalous Hall loop shows a strong PMA, a characteristic feature of Pt/FM based multilayers where interfacial spin-orbit coupling stabilizes the out-of-plane magnetization, as evidenced by a square out-of-plane hysteresis loop in Fig. 4b. To calculate the $H_k$, we apply an external field ($H_x$) parallel to the plane of the film and measured the Hall voltage. Fitting this behavior using the Stoner–Wohlfarth macrospin model yields a perpendicular anisotropy field $H_k \approx$ 4.9kOe which is consistent with expectations for multilayers exhibiting strong PMA. The angular dependence of the switching field exhibits an extended plateau, with $H_{sw}$ remaining nearly constant up to $\theta_H \approx$ 70°, followed by a steep increase at larger angles. This behavior indicates a further suppression of angular scaling compared to the single-stack structure and can be attributed to multilayer-induced modifications of the domain-wall depinning process. In the double stack, domain walls in adjacent CoFeB layers are coupled through magneto-static interactions, forming a composite wall with enhanced effective stiffness and a modified pinning landscape. The increased in-plane field component more efficiently reduces the effective depinning barrier by altering the internal structure of the coupled domain walls, allowing compensation for the reduced out-of-plane driving field over a wider angular range. As a result, the switching field remains weakly dependent on $\theta_H$ up to larger angles. Once the domain-wall structure reaches its equilibrium configuration, this softening effect saturates, and the continued reduction of the out-of-plane field component leads to a rapid increase in $H_{sw}$. The larger threshold angle value of $t = 70°$ obtained from the modified Kondorsky fit reflects this enhanced and prolonged compensation mechanism arising from interlayer coupling in this multilayer structure.

The DW velocity measurements were conducted quasi-statically using the MOKE microscope with the application of magnetic field pulses. Details of the measurement method are available in the literature[38]. After the nucleation process, five sequential magnetic pulses of a particular amplitude and pulse width were applied to calculate the average displacement. This process was repeated by changing the pulse duration 15 times to find the corresponding displacement v/s pulse width graph. To determine the velocity as a function of the magnetic field, approximately 3,400 domain images were captured and analyzed using an automated system[39]. Fig. 5a, b shows the p-MOKE hysteresis loop of both samples, confirming the presence of out-of-plane magnetic anisotropy. Coercivity $H_C$= 25.2 Oe for [Pt/CoFeB/Ru]$_{\times 2}$ sample is nearly double the $H_C$= 12.6 Oe of [Pt/CFB/Ru]$_{\times 1}$ because of increased pinning sites caused by increased interfaces. Magnetization reversal starts with bubble domain nucleation and expansion via DW motion along the magnetic easy axis. The increasing size of bubble domains after a particular nucleation pulse of amplitude 21.1 Oe and pulse width of 0.05 seconds for both samples, followed by subsequent five pulses of amplitudes 7.8 Oe and 18.7 Oe with pulse widths of 0.76 seconds and 0.46 seconds for [Pt/CFB/Ru]$_{\times 1}$ and [Pt/CoFeB/Ru]$_{\times 2}$, are shown in FIG. 5c and 5d respectively. Such bubble domains are observed in Pt/Co/Pt, Pt/Co/TiO$_2$/Pt[40]. R. S. Bevan et al.[41] also observed bubble domains in Si/SiO$_2$/Pt (5 nm)/CoFeB (0.7 nm)/Pt (3 nm) and explained bubble domains bounded by roughly circular domain walls, which are characteristic of high-quality films with strong PMA. Lower bubble domain density in such CoFeB system arises due to the weaker PMA as compared to the Co system. But in general, factors such as defects density can also affect the nucleation density. D.-Y. Kim at al.[40] also observed such bubble domains for $t_{Co}$< 1.0 nm in Ta (5.0 nm)/Pt (2.5 nm)/Co ($t_{Co}$ nm)/ TiO$_2$ (1.5 nm)/Pt (2.0 nm) films deposited on a Si/SiO$_2$(100 nm). The domination of DW energy over magnetostatic energy results in bubble domains formation. These bubbles act as nucleation spots branching into labyrinth domains[42].

Velocity measurements (Fig. 5e) are conducted in the creep regime, covering the domain wall speed slower than a few mm/sec. At respective coercive fields, DW velocities for [Pt/CoFeB/Ru]$_{\times 1}$ and [Pt/CFB/Ru]$_{\times 2}$ samples observed are 1142.58 µm/sec (1.14 mm/sec) and 1175.93 µm/sec (1.17 mm/sec) suggesting the DW velocity increase for increased interfaces. In the creep regime, the DW velocity follows the universal creep law as [43,44],

$$v = v_0 \exp\left(-\frac{\alpha}{k_B T} H^{-1/4}\right) \qquad (3)$$

where the pre-factor known as characteristic speed, $v_0 = \frac{\xi}{\tau}\exp\left(\frac{U_C}{k_B T}\right)$, creep scaling constant $\alpha = U_C H_{dep}^{\frac{1}{4}}$, $v$ and $H$ are the DW velocity and applied Magnetic field, $k_B$ and T are the Boltzmann's constant and temperature, $H_{dep}$ is the depinning field, $\xi$ is the correlation length of the disorder, $1/\tau$ is the attempt frequency at passing barriers in the wall motion[44]. The $U_c$ is related to the energy barrier separating one metastable state to another. For an elastic line i.e., the DW in our case, we have $U_c \propto \xi^2$ [44–46]. To investigate changes in creep regime parameters, we plotted DW velocity in linear form (Fig. 5f) using the equation:

$$\ln(v) = \ln(v_0) - \frac{\alpha}{k_B T} H^{-\frac{1}{4}} \qquad (4)$$

the linear plot of $\ln(v)$ vs $H^{-\frac{1}{4}}$ (Fig. 5f) is the confirmation of measurements covering only the creep regime. the value of $\alpha$ = 1.336 (±0.007) ×10$^{-20}$ JT$^{1/4}$ for [Pt/CoFeB/Ru]$_{\times 1}$ sample and 2.109 (±0.015) ×10$^{-20}$ JT$^{1/4}$ for [Pt/CoFeB/Ru]$_{\times 2}$ sample are obtained. A similar α value is also observed for Si/SiO2/Ta (5.0 nm)/Pt (1.5 nm)/Co (0.3 nm)/Pt (1.0 nm)[47]. As $\alpha \propto K^{5/8}$, K is an anisotropy constant[47], magnetic anisotropy strength increases with increasing interfaces. Under the limit of creep regime for [Pt/CoFeB/Ru]$_{\times 1}$ sample at H = 2.3 mT, ~18.7 times higher velocity is observed compared to [Pt/CoFeB/Ru]$_{\times 2}$ sample for the same H value.

## II.   Conclusions

In conclusion, we have studied magnetic and magneto-optical properties of Pt/CoFeB/Ru multilayer system which is expected to have large DMI and considerable perpendicular magnetic anisotropy. MOKE and AHE studies confirmed the out-of-plane magnetization of the films below a certain thickness of 1.2 nm. The effective magnetic anisotropy $K_{eff}$ of $0.88 \times 10^6$ erg/cm$^3$ has been achieved without annealing the system. The angular dependence of the switching field deviates from the conventional Kondorsky model and is well described using a modified Kondorsky formalism, capturing the role of field-induced domain-wall softening and pinning effects in the reversal process. Bubble domains formation in the MOKE microscopy are direct consequences of the PMA in the films. Polar magneto-optical Kerr microscopy observations of magnetic domains during the magnetization reversal process reveal detailed insights into the nucleation and movement of domain walls. Therefore, Pt/CoFeB/Ru can be useful as a promising candidate for future skyrmion-based device where domain wall control and magneto-optical response are required.

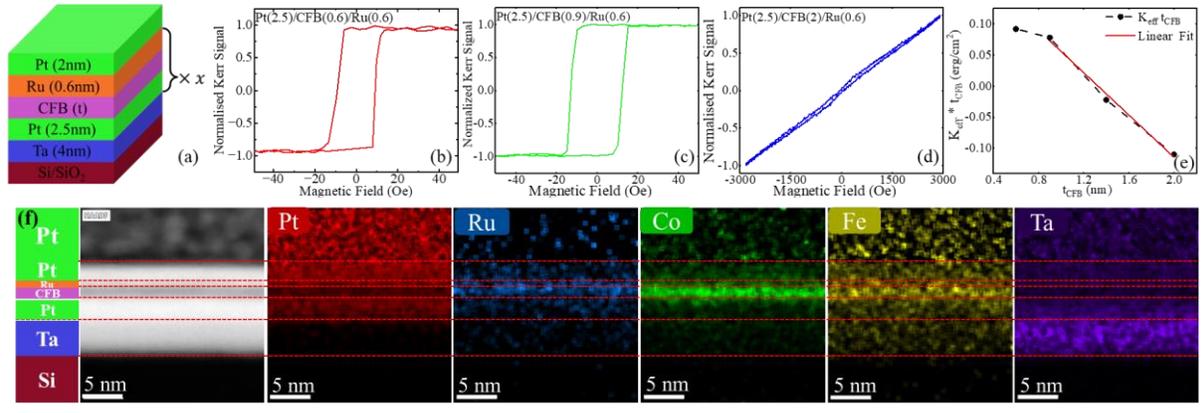

FIG 1: (a) Scheme of the deposited sample (a) Si/SiO$_2$/Ta(4)/Pt(2.5)/CoFeB(t)/Ru(0.6)/Pt(2). (b)-(d) Magnetic reversal loops of multilayers measured at room temperature with OOP MOKE. (e) Variation of $K_{eff} \times t_{CoFeB}$ with $t_{CoFeB}$ for Pt(2.5)/CoFeB(t)/Ru(0.6) thin films. (f) HAADF-STEM imaging combined with energy dispersive X-ray (EDS) spectroscopy with compositional mapping of elements, including Ta, Pt, CoFeB, Ru and near the interface.

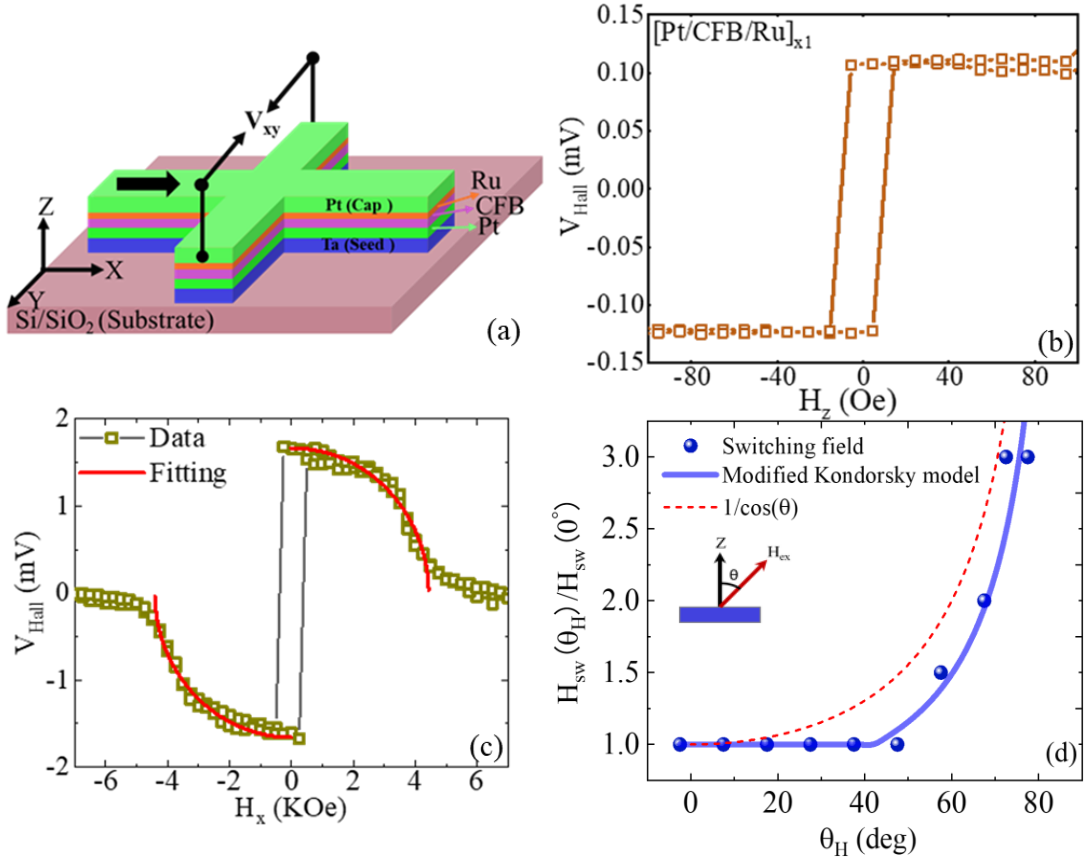

FIG. 2. (a) Schematic representation of the Hall bar structure and four-point Hall measurements based on the relative orientation of the magnetization ($M$), in-plane field ($H_x$), in-flow current ($J_c$), and transverse field ($H_z$). (b) Out-of-plane hysteresis loop measured by the anomalous Hall effect. (c) The evaluation of Hall voltage as a function of in-plane magnetic field ($H_x$). (d) $H_{sw}$ plots with respect to $\theta$. The inset represents the applied field with normal to film plane. The red line shows the $H_{sw}(\theta) = H_{sw}(0)/cos\theta$ curve, which is also known as the Kondorsky model[36] and the blue fitting is for the modified Kondorsky model.

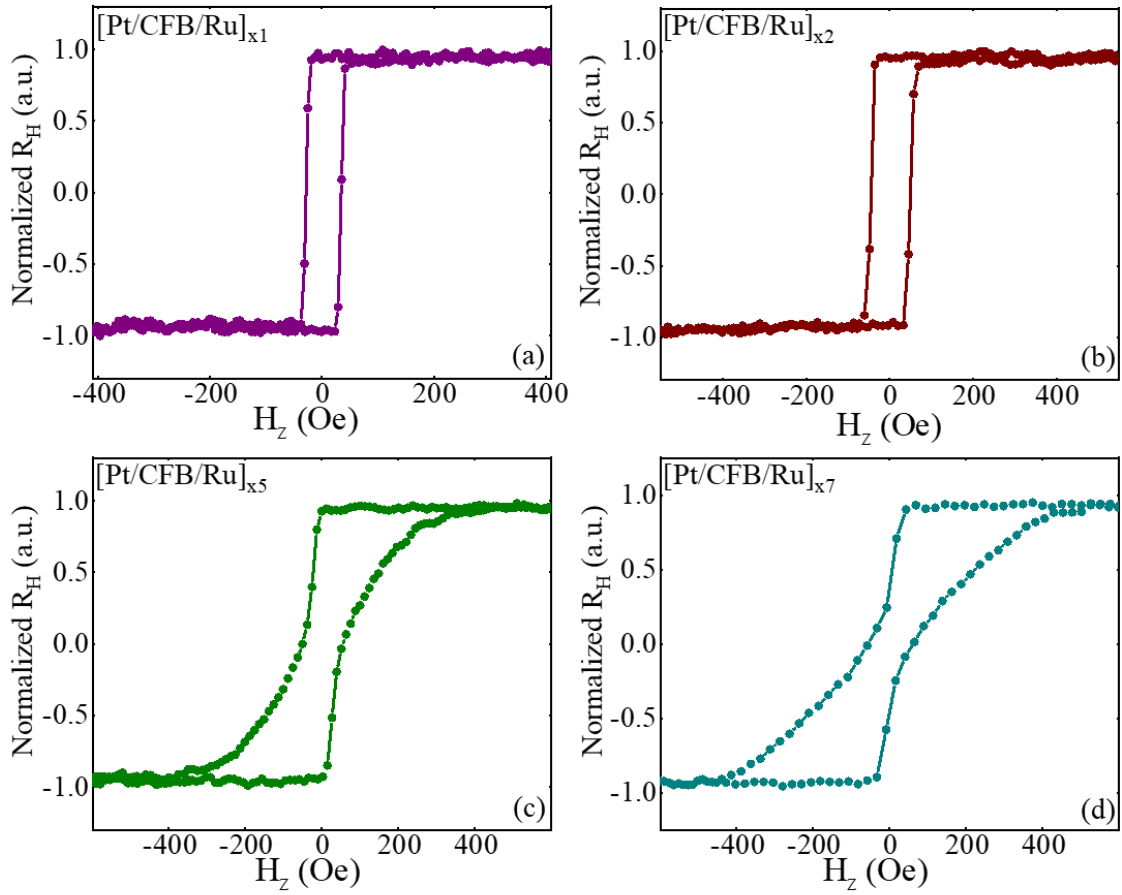

FIG 3: (a) AHE hysteresis loops as a function of out-of-plane magnetic field, $H_z$ for samples with different number stacks [Pt(2.5 nm)/CoFeB(0.9)/Ru(0.6 nm)]$_x$ where, (a) x=1, (b) x=2, and (c) x=5, (d) x=7.

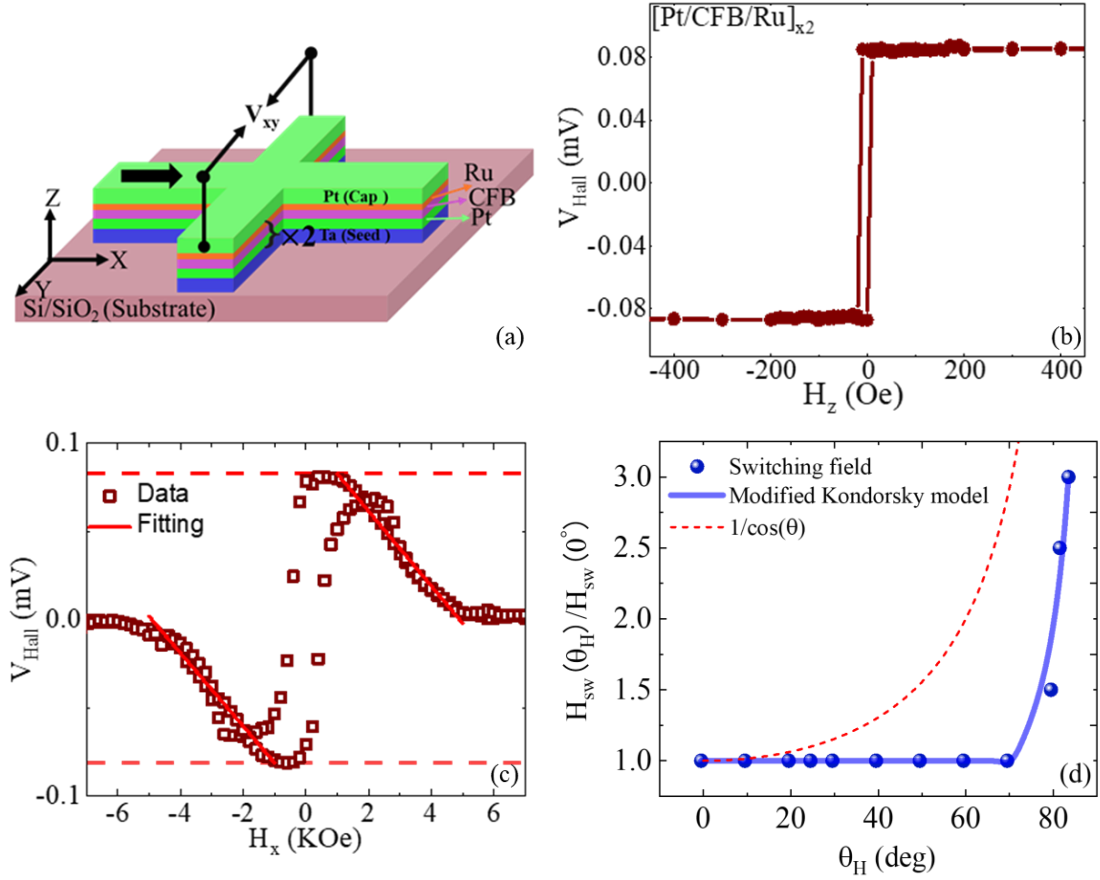

FIG 4: (a) The Schematic image of the Hall bar and the measurement configuration of anomalous Hall resistance. (b) Out-of-plane hysteresis loops for samples with 2 multilayer stacks. (c) Normalized Hall voltage as a function of in-plane applied magnetic field ($H_x$). (d) $H$sw plots as a function $\theta$. The red line shows the $H_{sw}(\theta) = H_{sw}(0)/cos\theta$ curve, which is known as the Kondorsky model and the blue line is for modified Kondorsky model.

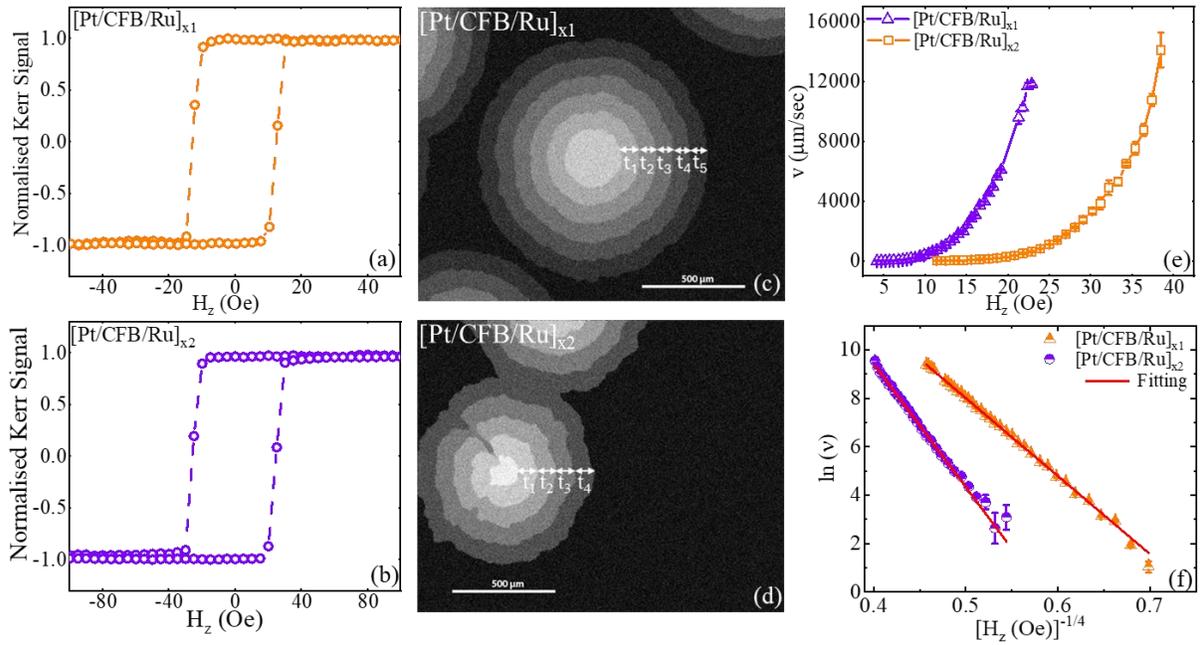

FIG 5: Square Hysteresis loops in OOP field direction for (a) [Pt/CoFeB/Ru]$_{\times 1}$ and (b) [Pt/CFB/Ru]$_{\times 2}$ samples confirm PMA. Bubble domain displacement for (c) [Pt/CoFeB/Ru]$_{\times 1}$, (d) [Pt/CFB/Ru]$_{\times 2}$ with series of field pulse amplitude of 7.8 Oe, 18.7 Oe, and pulse width of 0.76 seconds, 0.46 seconds, respectively. DW velocity measurements (e) v vs H plot and (f) linear plot of ln v and H$^{(-1/4)}$ in creep regime with fitting.